# QUANTUM BEHAVIOUR OF THE FLUX TUBE:

## A comparison between QFT predictions and lattice gauge theory simulations

F. GLIOZZI

*Dipartimento di Fisica Teorica dell' Università di Torino*
*Via P.Giuria 1, I-10125 Torino, Italy*

ABSTRACT

We review some universal features of the colour flux tube of gauge theories in the confining phase predicted by the infrared conformal limit of the underlying string theory. In particular we discuss shape effects in Wilson loops and rederive in a general way the logarithmic growth of the mean square width of the flux tube as a function of the interquark separation. Recent data on $3D$ $Z_2$ gauge theory, combined with high precision data on the interface physics of the $3D$ Ising model fit nicely to this behaviour over a range of more than two orders of magnitude.

## 1. Introduction

### 1.1. The String Picture of the Flux Tube

Experience on lattice gauge theory has shown that the relevant degrees of freedom of whatever gauge system in the confined phase are concentrated, in the infrared region, inside string-like flux tubes connecting among them the sources of the gauge field. This suggests, according to an old conjecture[1,2], that the vacuum expectation values of large Wilson loops $W(C)$, once the integration on the irrelevant degrees of freedom has been performed, are given by an effective string theory describing the dynamics of the flux tube as follows

$$\langle W(C) \rangle = \sum_{\{\Sigma: \partial\Sigma = C\}} e^{-S(\Sigma)} \quad , \tag{1}$$

where $\Sigma$ is a surface bounded by the loop $C$ and $S$ is a (largely unknown) string action. Within $SU(N)$ gauge theories there is an internal support of this idea: the $1/N$ expansion of weak coupling perturbation theory can be interpreted, according to 't Hooft analysis[2], as a topological expansion of a string theory in the number of handles of $\Sigma$ in which $1/N$ acts as the string coupling $g$. For large $N$ the vacuum expectation value $\langle W(C) \rangle$ is dominated by planar Feynman graphs. In the dual language of strings this implies that the sum in Eq. 1 is saturated by surfaces with no handles and the evaluation of $\langle W(C) \rangle$ is greatly simplified.

It has been recently shown [3] that in two-dimensional QCD on a Riemann surface the coefficients of the $1/N$ expansion have a simple, exact interpretation in terms of a string theory.

### 1.2. Condensation of Handles

For finite $N$, or for discrete gauge groups, the string enters a non-perturbative regime where the creation of handles becomes an important process. For this reason a stringy description of such gauge systems, which are the most interesting from the physical point of view, is often considered hopeless.

There is however an old argument [4], corroborated by recent numerical simulations [5,6], which suggests that this non-perturbative regime has actually a simple description. Indeed it has been shown long ago [4] that in critical string theory the sum of the divergent part of the handle insertions in any string amplitude can be reabsorbed in the renormalization of the string tension $\sigma$ and of the string coupling $g$. An indirect check of this property comes from the simplest gauge system, namely the three-dimensional $Z_2$ gauge model. Here, using the duality relation with the Ising model, it is possible to describe the properties of the effective string in terms of the fluid interfaces between domains of opposite magnetization. It is also possible to study the topological properties of these surfaces and it turns out[5] that the number of handles is very large and proportional to the area of the surface. Such a condensation of handles seems an ubiquitous phenomenon of the physics of random surfaces with fluctuating topology[6,7]; its meaning is rather simple: the mean handle is microscopic and its size $\ell$ does not depend on that of the surface. As a consequence, at a scale larger than $\ell$ there are very few visible handles, so the effective string coupling $g$, being proportional to the amplitude of handle formation, is vanishing, like the perturbative coupling in the large $N$ gauge systems.

The above argument strongly suggests that at large distances the functional integral given in Eq. 1 is dominated, like in the $1/N$ expansion, by surfaces with no handles, so the effective string description of the infrared region should greatly simplify.

### 1.3. $\alpha'$ Expansion and the Universality Class of the Flux Tube

Linear confinement implies that the action $S$ is proportional, in the infrared region, to the area of the surface $\Sigma$ swept by the flux tube, then the string action should be well approximated by the Nambu-Goto action, even if this simple string picture does not work at short distances [8].

A further simplification of the string action comes from the so called $\alpha'$ expansion, where $\alpha' = 1/2\pi\sigma$ : for large rectangular Wilson loops of size $L_1 \times L_2$ one can expand the action $S$ in the natural adimensional parameter $1/(\sigma L_1 L_2)$, obtaining

$$S = \sigma L_1 L_2 + \frac{1}{2} \int_0^{L_1} d\xi_1 \int_0^{L_2} d\xi_2 \sum_i^{D-2} (\nabla h_i(\xi_1, \xi_2))^2 + O\left(\frac{1}{\sigma L_1 L_2}\right) \quad , \tag{2}$$

where $h_i$ are the two-dimensional fields describing the transverse displacements of the flux tube. Thus the effective string theory reduces in the infrared limit to a massless gaussian model.

The above conclusion can be actually obtained also by more general considerations. Indeed strong coupling ($\beta$) expansions of $\langle W(C) \rangle$ in whatever gauge theory show that the flux tube undergoes a transition towards a rough phase for $\beta$ larger than a given $\beta_r$ (the roughening point). It is widely believed that this transition is described by the universality class of Kosterlitz and Thouless[9] (KT). Thus the effective string action, even if it is substantially unknown, should belong to the same universality class of the $XY$ model. The renormalization group equations of the KT universality class can be simply expressed in terms of two couplings:

$$\dot{x}(t) = -y(t)^2 \quad , \tag{3}$$
$$\dot{y}(t) = -x(t)y(t) \quad , \tag{4}$$

where $t$ is the RG scale parameter, $x = \pi T - 2$, ($T$ is the temperature of the $XY$ model) and $y$ is the vortex fugacity. The inverse temperature can be identified in the infrared limit with the surface stiffness $\kappa$, which in the continuum limit coincides with the string tension. These RG equations describe the transition from a smooth phase (with free vortices) to a rough phase where the effective vortex fugacity flows to zero and then the the theory becomes at large scale a massless free field theory.

So, we have reached a conclusion similar, but more general than that obtained by the $\alpha'$ expansion of the Nambu-Goto action: for large enough Wilson loops it is not necessary to know the specific form of the string action because it flows towards a free massless theory. This limit action is of course critical in the whole rough phase, then it produces universal (i.e. gauge group independent) finite size effects. Moreover, because this theory is also infrared divergent, such finite size effects are expected to be rather strong. We shall discuss two of these universal effects: the shape dependence of the Wilson loops and the growth of the mean width of the flux tube.

## 2. Shape Effects in the Wilson Loops

### 2.1. The Functional Form of the Quantum Fluctuations

The contribution of the first few terms of the $\alpha'$ expansion to the vacuum expectation values of the Wilson loops can be calculated exactly[10]. Here we consider only the contribution of the gaussian term, which is the one associated to the universal effects.

For large enough Wilson loops we can parametrize $\langle W(L_1, L_2) \rangle$ as follows

$$-\log\langle W(L_1, L_2)\rangle = \sigma L_1 L_2 + p(L_1 + L_2) + c + q(L_1, L_2) + O(\alpha') \tag{5}$$

where the string tension $\sigma$ and the the two parameters $c$ and $p$ are the only free parameters of this equation; $q(L_1, L_2)$ is the universal contribution due to the gaussian functional integral and is a know function of the shape $(L_1, L_2)$ which does not depend on the gauge system, but only on the choice of the boundary conditions. For instance, for fixed boundary conditions we have

$$q(x, y) = \frac{D-2}{2} \log\left(\frac{\eta(iy/x)}{\sqrt{x}}\right) \quad , \tag{6}$$

where the Dededekind eta function is given by

$$\eta(\tau) = q^{\frac{1}{24}} \prod_{n=1}^{\infty} (1 - q^n) \quad , q = e^{2\pi i \tau} \quad . \tag{7}$$

The older fits of Wilson loop data put arbitrarily $q(x,y) = 0$ with unacceptable $\chi^2$ values. Insertion of the quantum correction given in Eq. 6 improves considerably the fits, even if discrepancies have been observed for various gauge systems both in $D = 3$ and $D = 4$ dimensions [11,12].

It has been suggested that these discrepancies might be due to the choice of fixed b.c.[13]. Actually the gaussian model is a member of a one-parameter family of conformal field theories ( those with central charge $c = 1$) having the same local behaviour, but different topological modes (winding modes around a circle of radius $r$ which is the parameter of this family). If the free theory describing the infrared limit of the effective string is allowed to have such topological modes, it is possible to introduce new kinds of boundary conditions which give a much better description of the Wilson loop. Roughly speaking, such a topological modification of the gaussian model allows one to treat in a different manner quarks line and antiquarks, while in the ordinary gaussian model there is no way to distinguish a flux tube connecting two quarks from that connecting a quark-antiquark pair, which from the point of view of the physics of the gauge system are of course completely different.

2.2. Fluid Interfaces

In order to test more accurately the above universal asymptotic behaviour it would be convenient to study the simplest gauge system, i.e. the $3D$ $Z_2$ gauge model. It is dual to the $3D$ Ising model where one could use non-local cluster algorithms for numerical simulations which yield a drastic reduction of the statistical error compared with the old Metropolis or Heat Bath methods. Direct measurements of expectation values of the Wilson loops with this new technique have not yet been performed, however there is a related observable that has been evaluated in this way. Let us see briefly the method.

Denote by $\sigma_l = \pm 1$ the link variables of the $Z_2$ gauge system and by $\tilde{\sigma} = \pm 1$ the site variables of the dual Ising model. One can relate the expectation value of the plaquette $P = \sigma_1 \sigma_2 \sigma_3 \sigma_4$ in terms of dual variables as follows

$$\langle P \rangle_{gauge} = \langle \exp(-2\tilde{\beta}\tilde{\sigma}_l) \rangle_{spin} \quad , \tag{8}$$

with $\tilde{\sigma}_l = \tilde{\sigma}_1 \tilde{\sigma}_2$, where $l$ is the link of the dual lattice orthogonal to the plaquette $P$ and punching the center of the plaquette; $\tilde{\beta}$ is related to $\beta$ by

$$\tilde{\beta} = -\frac{1}{2} \log (\tanh \beta) \quad . \tag{9}$$

More generally, denoting by $\Sigma$ an arbitrary surface with boundary $\partial \Sigma$, we have

$$\langle W(\partial \Sigma) \rangle_{gauge} = \langle \prod_{l \in \Sigma} \exp(-2\tilde{\beta}\tilde{\sigma}_l) \rangle_{spin} \quad , \tag{10}$$

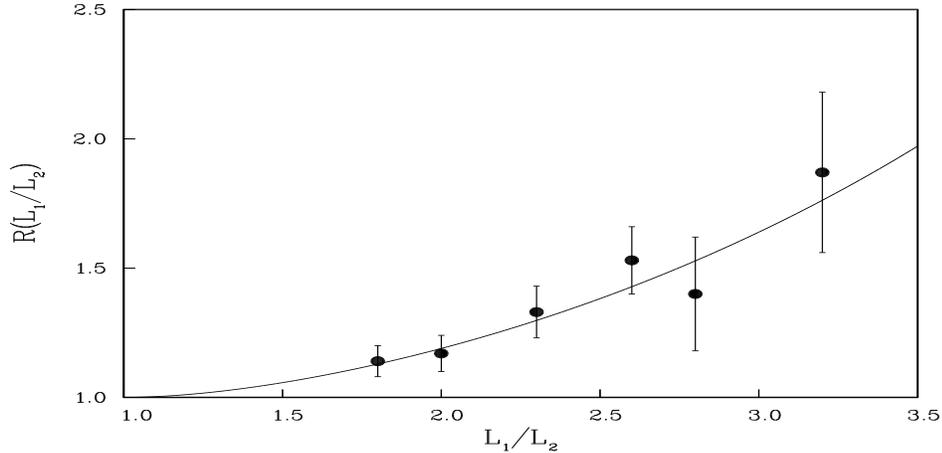

Fig. 1. Finite size effects in fluid interfaces.

where the product is over all the dual links cutting the surface $\Sigma$. In the dual Ising model with periodic b.c. in $x$ and $y$ directions one can also realize a sort of Wilson loop which has no analogue in the gauge system, by choosing as $\Sigma$ a toroidal surface wrapped around the periodic directions ( for instance a plane orthogonal to the $z$ axis). Then $\langle \prod_{l \in \Sigma} \exp(-2\tilde{\beta}\tilde{\sigma}_l) \rangle$ denotes the interface partition function. This observable is particularly well suited to study the shape dependent effects, because there are no complications due to the boundary sources. The size of this periodic Wilson loop is four times larger than the maximal ordinary loop one can measure in the corresponding gauge system. The functional form of Eq. 5 is now modified because the perimeter term is absent, of course, and the the gaussian term is slightly different with respect to Eq. 6 owing to the periodic boundary conditions. One gets

$$q(x,y) = \log\left(\sqrt{\Im m \tau}|\eta(\tau)|^2\right) \quad , \tau = i\frac{x}{y} \quad . \tag{11}$$

A set of numerical simulations on the Ising model[14] and on the three-state Potts model[15] gave a nice, accurate confirmation of this behaviour for a wide range of $\tau$.

In particular, the ratio $R = Z(L_1, L_2)/Z(\tilde{L}, \tilde{L})$ of interfacial partition functions for systems with the same area $(L_1 L_2 = \tilde{L}^2)$, but different shape, is only a function of $L1/L2$ and is directly expressible in terms of Eq. 11 with no free parameters. In Fig.1 a comparison is made between such theoretical prediction and a set of numerical simulations in the Ising model at $\beta = .2275$ (black dots).

## 3. Width of the Colour Flux Tube

### 3.1. Thickness of the Flux Tube as an Effect of String Fluctuations

One of the simplest and general consequences of the existence of a thin flux tube connecting a quark pair in the confining phase is that this flux tube fluctuates. The effect of these quantum fluctuations is to produce an effective square width of the flux tube which grows logarithmically with the interquark distance.

Such a logarithmic behaviour has been predicted many years ago by Lüscher, Münster and Weisz [16] in the framework of the effective string picture of the gauge systems. In the next subsection we shall rederive and refine this universal law by directly using some exact results on two dimensional free field theory in a finite box.

Let us begin by defining the observable we want to discuss. The square width (or the square gyration radius) of the flux tube generated by a planar Wilson loop $W(C)$ is defined as the sum of the mean square deviation of the transverse coordinates of the underlying string, i. e.

$$w^2 = \frac{1}{A} \sum_{i=1}^{D-2} \int_{\mathcal{D}} d^2\xi \langle \left( h_i(\xi_1, \xi_2) - h_i^{CM} \right)^2 \rangle \quad , \tag{12}$$

where $\mathcal{D}$ is the planar domain bounded by $C$, $A$ its area $A = \int_{\mathcal{D}} d^2\xi$ and $h_i^{CM}$ is the transverse coordinate of the center of mass of the flux tube, given by

$$h_i^{CM} = \frac{1}{A} \int_{\mathcal{D}} d^2\xi \, h_i(\xi_1, \xi_2) \quad . \tag{13}$$

Combining Eq. 12 with Eq. 13 we get

$$\begin{aligned} w^2 &= \frac{1}{2(A)^2} \sum_{i=1}^{D-2} \int_{\mathcal{D}} d^2\xi \int_{\mathcal{D}} d^2\xi' \langle (h_i(\xi) - h_i(\xi'))^2 \rangle \\ &= \frac{1}{\sigma(A)^2} \sum_{i=1}^{D-2} \int_{\mathcal{D}} d^2\xi \int_{\mathcal{D}} d^2\xi' \left( G_i(\xi, \xi') - G_i(\xi, \xi) \right) \quad , \end{aligned} \tag{14}$$

where $G_i(\xi, \xi')$ is the Green function of the field $h_i$ defined as

$$G_i(\xi, \xi') = \frac{1}{\sigma} \langle h_i(\xi) h_i(\xi') \rangle \quad . \tag{15}$$

According to the discussion of §2, we expect that the 2D field theory describing the dynamics of the flux tube flows, for a domain $\mathcal{D}$ big enough, to a massless gaussian limit, where the Green function fulfills the free field equation

$$-\Delta G(\xi, \xi') = \delta^{(2)}(\xi - \xi') \quad . \tag{16}$$

For simplicity we omitted the transverse index $i$. It is convenient to adopt complex coordinates by defining $z = \xi_1 + i\xi_2$; then the solution $G_\infty(z, z')$ of the above equation in the infinite plane has the well known form

$$G_\infty(z, z') = -\frac{1}{2\pi} \log |z - z'| + c \quad , \tag{17}$$

where $c$ is an arbitrary constant. Note that for $z \sim z'$ the free Green function is ultraviolet divergent , then Eq. 14 is ill-defined and needs regularization. Using for instance the point-splitting method one gets at once for a square Wilson loop $L \times L$

$$w^2 = \frac{1}{2\pi\sigma} \log \left( L/R_c \right) \quad , \tag{18}$$

where the UV cut-off has been absorbed in the definition of the scale $R_c$.

## 3.2. Green Functions in a Finite Box

Actually the above calculation has a flaw, because the theory is only defined in a finite box, while it is used the propagator of the infinite plane. The Green function on a box has a very different form which depends on the shape of the box and on the choice of the boundary conditions. This fact might in principle modify the functional form of the width. Actually we shall see that this is not the case.

The problem of finding the free Green function on an arbitrary, simply connected region $\mathcal{D}$ of the plane with fixed boundary conditions has been already solved by the mathematicians of the last century. Accordingly, it can be reformulated in a way where $z$ and $z'$ play an apparently asymmetric role: find a real function $f_{z'}(z, \mathcal{D})$ which is harmonic in the punctured set $\mathcal{D} \setminus \{z'\}$, vanishes at the boundary $\partial \mathcal{D}$ and diverges logarithmically as $-\log|z - z'|$ for $z \to z'$. Such a problem can be solved once it is found a conformal mapping $z \to \ell$ of $\mathcal{D}$ onto the unit circle $|\ell| = 1$ which maps $z'$ into the origin $\ell = 0$. Denoting by $\ell_{z'}(z)$ the analytic function providing us with such a mapping, it is immediate to verify that

$$f_{z'}(z, \mathcal{D}) = \log|\ell_{z'}(z)| \quad , \tag{19}$$

and that the function

$$G_{\mathcal{D}}(z, z') = \frac{1}{2\pi} f_{z'}(z, \mathcal{D}) \tag{20}$$

satisfies Eq. 16. For, note that $G_{\mathcal{D}}$ is the real part of an analytic function, then $\Delta G_{\mathcal{D}} = 0$ for $z \neq z'$; since $z'$ is mapped into the origin $\ell = 0$, then $G_{\mathcal{D}}$ diverges logarithmically for $z \to z'$ and is normalized as required by the delta function; finally, being the boundary $\partial \mathcal{D}$ mapped into the unit circumference, $G_{\mathcal{D}}$ vanishes at this boundary as it should.

For our purposes, the relevant property of the conformal mapping $\ell$ is that one can perform an arbitrary scale transformation $z \to \lambda z$ without destroying the conformal character of $\ell$. More precisely we can write

$$f_{z'}(z, \mathcal{D}) = f_{\lambda z'}(\lambda z, \mathcal{D}_\lambda) \quad , \tag{21}$$

where $\mathcal{D}_\lambda$ denotes the scaled domain. A direct consequence is that it is always possible, for any finite $\mathcal{D}$, to fix the area of the scaled domain $\mathcal{D}_\lambda$ to an arbitrary value, say 1, without changing the Green function. It follows that the integration of the finite part $G(z, z')$ in Eq. 14 cannot depend on the size of the domain $\mathcal{D}$ but only on its shape. Surprisingly enough, the logarithmic growth of the square width $w^2$ comes from the UV divergent part. Let us see why. According to the point spitting method, we regularize $G(\xi, \xi)$ with the replacement

$$G(\xi, \xi) \to G(\xi, \xi + \varepsilon) \quad , \tag{22}$$

where $\varepsilon$ is the ultraviolet cut-off. Combining the logarithmic divergence of $G$ with its scaling property expressed in Eq. 21, we can write

$$G(\xi', \xi' + \varepsilon) \sim -\frac{1}{2\pi} \log(\varepsilon/L) + c \quad, \tag{23}$$

where $L$ is a typical linear dimension of the domain $\mathcal{D}$. Inserting such an expression in Eq. 14 we get the logarithmic law of Eq. 18 where now $R_c$ is a calculable function of the shape of the domain.

The physical meaning of such a behaviour is now clear: the dynamics of the flux tube is described by a truly free-field theory only at large distances; the cut-off $\varepsilon$ sets up the ultraviolet scale $R_c$ below which the free-field approximation breaks down. Obviously this scale cannot depend on the infrared scale $L$, then a variation of $L$ cannot be balanced by a variation of $\varepsilon$. In other terms the ultraviolet divergence breaks the scale invariance of the problem and then a variation of $L$ gives rise to an observable effect.

The volunterous reader may check the above arguments in a particularly simple example, given by the Green function on a disc of radius $L$

$$G_{disc}(z, z') = -\frac{1}{2\pi} \log \left| L \frac{z - z'}{L^2 - \bar{z}'z} \right| \quad . \tag{24}$$

For a rectangular box of size $L_1 \times L_2$, which is the most interesting case for the lattice gauge models, we have

$$G_{L_1, L_2}(z, z') = -\frac{1}{2\pi} \log \left| \frac{\sigma(z - z')\sigma(z + z')}{\sigma(z - \bar{z}')\sigma(z + \bar{z}')} \right| \quad , \tag{25}$$

where $\sigma(z)$ is the Weierstrass sigma function [17] for the rectangle of sides $2L_1$ and $2L_2$. Inserting this expression in Eq. 14 it turns out that the scale $R_c$ does not depend very much on the modulus $\tau = iL_2/L_1$ for not too elongated rectangles.

*3.3. Numerical Simulations*

Though the logarithmic growth of the squared width of the flux tube is the most important and model-independent quantum effect predicted by the effective string description, it has not yet been observed until now. This is due to two main reasons. First, this effect is an infrared phenomenon that can be seen, as we have explained, only for very large Wilson loops, which are at the boundary of the sizes reached by the present numerical simulations. Secondly, a logarithmic growth can be checked numerically with reasonable confidence only if one can deal with a set of very precise data in a wide range of the loop sizes, which is again a condition which the current lattice simulations for $SU(2)$ and $SU(3)$ gauge theories have not yet attained.

It has been recently observed[18] that in the simplest gauge system, i.e. the $Z_2$ 3D gauge model, it is nowadays possible to overcome these difficulties by exploiting the duality relation between this model and the ordinary 3D Ising model explained in §2.2 .

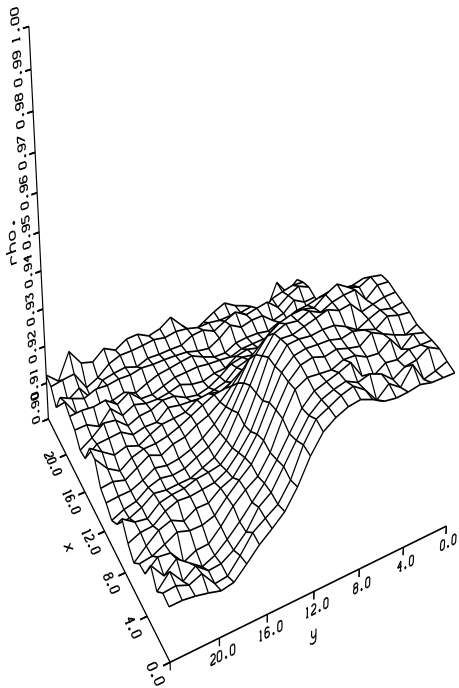

Fig. 2. Density of a $Z_2$ flux tube evaluated at a distance of two lattice spacings from a Wilson loop $24 \times 24$ sitting on the $xy$ plane. Only the upper left quadrant is reported

Indeed, using the one-to-one mapping of physical observables of the gauge system into the corresponding spin observables it is possible to replace the ordinary Metropolis or heath-bath method with a non-local cluster algorithm (for instance the Swendsen and Wang method). This allows us to probe the structure of the flux tube with an unprecedented accuracy. The procedure is the following. A Wilson loop $W(C)$ is realized in the spin lattice by frustrating all the links cutting a given surface $\Sigma$ bounded by $C$. These frustrated links modify the vacuum state so that the vacuum expectation value $\langle P \rangle_W$ of the plaquette, or better its spin analogue, is no longer translational invariant, but becomes a function of its relative position with respect to $W(L_1, L_2)$ and is related to the expectation value in the ordinary vacuum by

$$\langle P \rangle_W = \langle W(L_1, L_2) P \rangle / \langle W(L_1, L_2) \rangle \quad . \tag{26}$$

The difference between the expectation value of the plaquette in the vacuum modified by the presence of $W(L_1, L_2)$ and the expectation value in the ordinary vacuum can be considered as a measure of the density of the flux tube. Choosing for instance as a probe a plaquette $P_{\parallel}$ parallel to the plane of the Wilson loop we can write

$$\rho_{\parallel}(x, y, z) = \langle P_{\parallel} \rangle_W - \langle P \rangle \quad . \tag{27}$$

Other orientations of the plaquette give approximately the same distribution.[*] The flux density $\rho_{\parallel}$ can be viewed, up to an obvious normalization, as the probability for unit of volume of finding the fluctuating string at the point of coordinates $(x, y, z)$. The main difference between the flux density in $Z_2$ gauge model and that in theories with a continuous gauge group is that in the discrete case there is no self-energy peak associated to the quark lines ; as a consequence, the formation of the flux tube can be seen more neatly. Starting from a point on the perimeter of the Wilson loop an moving inwards, one finds that the flux density increases rapidly in a few lattice spacings and then it reaches a stable plateau. The same feature can be found in any plane parallel to the Wilson loop, with a value of the plateau which is only a function of the distance from the plane of the loop as Fig. 2 shows.

---

[*]This seems not to be the case in four dimensional systems[19,20]

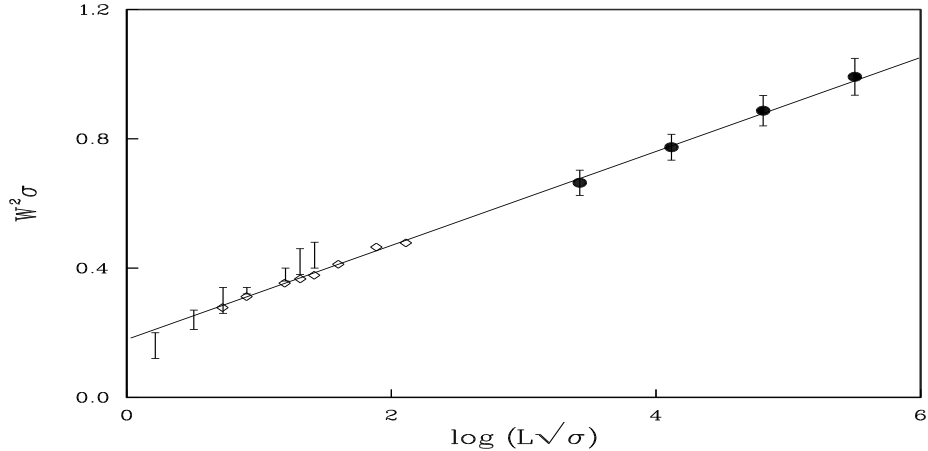

Fig. 3. Squared width of the flux tube in units of sigma.

Then, assuming that the Wilson loop is located in the plane $z = 0$, the mean square width of the flux tube can be defined as

$$w^2 = \int d\,x^3\, z^2 \rho_{\|}(x,y,z) \,/\, \int d\,x^3 \rho_{\|}(x,y,z) \quad , \qquad (28)$$

where the integration volume is bounded by the plateau described above. We performed two sets of simulations. The first was at $\beta = 0.7516$ corresponding to[†] $\sigma a^2 = 0.0107(1)$ . We considered six different squared Wilson loops of sides ranging from $L = 16$ to $L = 40$ and fitted the data to the two-parameter formula

$$w^2 = a\,\log(L) + b \quad . \qquad (29)$$

We got $b = -17.4 \pm 4.0$ and $a = 14.4 \pm 1.2$. The theoretical value $a_{th}$ of the parameter $a$, fixed by Eq. 18 to be $a_{th} = \frac{1}{2\pi\sigma}$ , matches nicely with this value, indeed we have $a_{th} = 14.8$ . The second set of simulations was made at $\beta = 0.7460$ corresponding to $\sigma a^2 = 0.0189(1)$ with squared Wilson loops of sides ranging from $L = 15$ to $L = 60$. Fits to Eq. 29 gave $b = -5.8 \pm 2.0$ and $a = 7.7 \pm 0.6$ while the theoretical value is $a_{th} = 8.42$ .

---

[†]The high precision of this value is a consequence of the use of the non-local cluster algorithm.

An internal consistency check of these numerical data comes from a comparison between Eq. 18 and Eq. 29. We get that the physical adimensional quantity $\sqrt{\sigma}R_c$ is expressed in terms of $a$ and $b$ as

$$\sqrt{\sigma}R_c = \frac{\exp -b/a}{\sqrt{2\pi a}} \quad . \tag{30}$$

Using the fitted values of the parameters, we get $\sqrt{\sigma}R_c = 0.35 \pm 0.11$ at $\beta = 0.7516$ and $\sqrt{\sigma}R_c = 0.31 \pm 0.09$ at $\beta = 0.7560$ in good agreement with scaling.

One of the important features of Eq. 18 is that it can be written in a universal form by expressing all the dimensional quantities $w$, $L$ and $R_c$ in units of $\sqrt{\sigma}$. Accordingly we report in Fig. 3 data from different $\beta$'s and and also recent very accurate data[21] for the mean square width of fluid interfaces in the dual Ising model (black dots). The straight line represents the logarithmic fit to our data at $\beta = 0.7460$ (rhombs). Within the statistical accuracy these data clearly support a logarithmic widening of the flux tube in a range of quark separation $L$ of more than two orders of magnitude, starting about $L_{min}\sqrt{\sigma} \simeq 1.7$.

In order to compare these results with analogous data for other gauge groups, we can, with an abuse of language, express $\sqrt{\sigma}$ in the same physical units of the QCD string . Then the crossover to the infrared logarithmic behaviour observed in Fig. 3 corresponds to $L_{min} = .75$ fm while the maximal probed elongation corresponds to more than 100 fm.

The data on $4D$ $SU(2)$ flux tube[19,20] cover now a distance up to 2 fm, but are still affected by strong systematic errors and are compatible also with a constant width for separations larger than 1 fm. They are scattered mainly above the straight line of Fig. 3, suggesting a value of the UV scale a bit smaller than the value of the $Z_2$ system, while the crossover distance is approximately the same.

Such a similarity of the cross-over distance $L_{min}$ in so different gauge systems has a plausible explanation. According to Eq. 2, $L_{min}$ should be controlled by the first non gaussian term in the $\alpha'$ expansion the effective string action. This term does not depend on the nature of the gauge group, so it should produce universal effects. Actually it has been recently[22] shown that this term describes accurately some finite size effects in fluid interfaces which cannot explained in the free field limit.